\documentclass[runningheads]{llncs}
\usepackage[T1]{fontenc}
\usepackage{amsfonts}
\usepackage{amsmath} 
\usepackage{graphicx}
\usepackage{xcolor}
\usepackage{multirow}
\usepackage{hyperref}
\begin{document}
%
\title{BioSonix: Can Physics-based Sonification Perceptualize Tissue
Deformations From Tool Interactions?}

\titlerunning{BioSonix}

\author{Veronica Ruozzi*\inst{1}
\and
Sasan Matinfar*\inst{2,4}
\and
Laura Schütz\inst{2}
\and
Benedikt Wiestler\inst{3,4}
\and
Alberto Redaelli\inst{1}
\and
Emiliano Votta\inst{1}
\and
Nassir Navab\inst{2,4}
}

\authorrunning{Ruozzi, Matinfar et al.}

\institute{Department of Electronics Information and Bioengineering, Politecnico Di Milano, Italy\\
\and
Computer Aided Medical Procedures, Technische Universität München, Germany\\
\and 
Neuroradiology Department, Klinikum Rechts der Isar, TU Munich, Germany\\
\and 
Munich Center for Machine Learning (MCML)\\
\email{veronica.ruozzi@polimi.it}\\
}

\maketitle              
\begin{abstract}
Perceptualizing tool interactions with deformable structures in surgical procedures remains challenging, as unimodal visualization techniques often fail to capture the complexity of these interactions due to constraints such as occlusion and limited depth perception. This paper presents a novel approach to augment tool navigation in mixed reality environments by providing auditory representations of tool-tissue dynamics, particularly for interactions with soft tissue. BioSonix, a physics-informed design framework, utilizes tissue displacements in 3D space to compute excitation forces for a sound model encoding tissue properties such as stiffness and density. Biomechanical simulations were employed to model particle displacements resulting from tool-tissue interactions, establishing a robust foundation for the method. An optimization approach was used to define configurations for capturing diverse interaction scenarios with varying tool trajectories. Experiments were conducted to validate the accuracy of the sound-displacement mappings. Additionally, two user studies were performed: the first involved two clinical professionals (a neuroradiologist and a cardiologist), who confirmed the method’s impact and achieved high task accuracy; the second included 22 biomedical experts, who demonstrated high discrimination accuracy in tissue differentiation and targeting tasks. The results revealed a strong correlation between tool-tissue dynamics and their corresponding auditory profiles, highlighting the potential of these sound representations to enhance the intuitive understanding of complex interactions.


\text{*} contributed equally to this work.

\end{abstract}

\section{Introduction}
Intraoperative decision-making relies heavily on medical imaging; however, interpreting these images is influenced by cognitive biases, as noted in~\cite{chen2023investigating}. The challenges are heightened in percutaneous and minimally invasive surgeries, where constrained access, lack of haptic feedback, and reliance on two-dimensional imaging place significant cognitive demands on surgeons. These limitations hinder spatial reasoning, instrument localization, and dynamic anatomical mapping. Robot-assisted surgeries add to the difficulty by introducing trust issues with the system. The complexity intensifies during soft tissue manipulation, such as tumor resection, where tissue deformations deviate from preoperative plans, complicating real-time awareness and precise instrument placement within narrow, deformable pathways. Medical Augmented Reality Systems (MARS) have shown great promise in improving preoperative planning, real-time visualization, and intervention precision. By augmenting patient anatomy, tool pathways, and catheter positions, MARS enhances spatial understanding of tool-target relationships. Its effectiveness in image-guided interventions is evident in orthopedics~\cite{ORTHO-XR-REVIEW}, brain surgery~\cite{BRAIN-XR}, and other areas such as pulmonary and abdominal procedures~\cite{PULMONARY-XR-REVIEW,ABDOMINAL-XR}.

However, unimodal visual channels impose limitations, including occlusion, depth perception challenges, and frequent shifts of focus between displays, complicating hand-eye coordination. Humans excel at solving complex problems and performing intricate tasks through multisensory processing~\cite{shams2008benefits,ernst2011multisensory}. By integrating visual, tactile, and auditory inputs, they form a holistic understanding of their environment, shaped by technical skills, experience, and intuition. However, current MARS systems fail to fully harness this capability, emphasizing the need for advanced cognitive support systems that bridge human-machine interaction. Delivering real-time, context-sensitive feedback aligned with human perception can enhance trust, augment cognitive abilities, improve precision, and empower surgeons to navigate complex scenarios with greater confidence.

\subsection{Challenges in Soft Tissue Modeling and Augmentation}
Current systems address soft tissue challenges through two primary strategies: modeling deformations accurately and augmenting these models effectively for the surgeon.

\textbf{Soft Tissue Deformation Modeling –} A major challenge in AR technologies is registering preoperative imaging with the surgical field, made complex by patient positioning and soft tissue deformations. While AR excels in rigid-body contexts like orthopedics, where fixed structures such as bones or the skull facilitate alignment, soft tissue applications face greater difficulties. For example, brain shifts during craniotomies or tissue configuration changes due to patient positioning complicate registration and tool accuracy ~\cite{REVIEW-AR-VASCULAR}. Nonetheless, AR has shown promise in guiding soft tissue procedures like tumor ablation. By employing deformable registration techniques and integrating intraoperative imaging for real-time updates, AR systems can adapt to dynamic changes and improve surgical outcomes ~\cite{REVIEW-AR-TUMOR,Neuro-AR-navigation}.

Most AR systems depend on preoperative imaging, limiting adaptability to real-time intraoperative changes. While some integrate intraoperative data for navigation~\cite{Neuro-AR-navigation}, such cases are uncommon. The lack of dynamic adaptation to anatomical changes, especially those caused by tool navigation, is a significant limitation. This hinders the clinical adoption of AR systems, where real-time anatomical updates and precise targeting are crucial for successful minimally invasive procedures.


\textbf{Perceptual Augmentation --} Even with accurate deformation modeling, MARS faces the challenge of delivering intuitive, real-time feedback to surgeons and achieving seamless integration into the surgical workflow.

Haptic feedback, while essential for tool guidance, faces challenges such as integration issues, safety concerns with bi-directional systems, and replicating tactile sensations. Adoption is further limited by high costs, latency, and operating room constraints~\cite{HAPTIC-REVIEW-2020}.

While the human visual system excels at spatial representation, the auditory system efficiently captures dynamic object characteristics with exceptional temporal resolution. By perceiving sonic qualities like pitch, timbre, and rhythm, it enables rapid comprehension of complex data and swift responses to changes, making auditory cues an intuitive and cost-effective medium for surgical tasks. This has led studies~\cite{illanes2018novel,ostler2024sound} to explore real-time acoustic sensing for tool-tissue interactions, though the signals often remain imperceptible to humans.

\textbf{Sonification}, the transformation of data into sound, has proven effective for surgical applications by helping surgeons maintain focus, reduce distractions, and optimize outcomes~\cite{black2017survey,matinfar2023sonification,matinfar2018surgical,matinfar2019sonification,ziemer2023three,schuetz2024shape}. Methods for transforming basic data inputs, like multistate signals~\cite{matinfar2017surgical} or spatial mapping~\cite{matinfar2023sonification,ziemer2023three}, have proven effective in simple scenarios. However, these approaches are often prescriptive and fail to capture the complexity or provide detailed, textural insights into tool-tissue interactions. The dynamic nature of surgery—shaped by patient-specific anatomy, procedural variability, and diverse strategies—poses challenges for standardizing augmentations. Despite AI and robotics advances, human expertise remains unmatched in integrating sensory inputs and adapting to complexities. Augmentation systems must meet the nuanced demands of surgery to effectively support highly trained surgeons.

Sonification of high-dimensional medical imaging has been explored to preserve data richness~\cite{hermann2000sonification,gionfrida2017novel}, but usability challenges often limit accessibility, especially for non-musical users. The human auditory system efficiently processes spectro-temporal sound features within fractions of a second, shaping cognitive patterns~\cite{theunissen2014neural,moerel2012processing}. Subtle sound differences are easily perceived, with those resembling real-world auditory experiences often felt as more intuitive. The effectiveness of these sounds relies on the relationship between sound and gesture, as well as users’ prior experiences with everyday objects~\cite{chase2023realism,susini2012naturalness,lemaitre2009toward}. A study proposed model-based sonification as an alternative to simple parametric mapping, transforming high-dimensional medical imaging into physics-driven sound models to intuitively represent tissue textures, with stiffer tissues sounding distinct from softer ones~\cite{matinfar2023tissue}. This approach showed promise in interactive scenarios like retinal surgery~\cite{matinfar2024ocular} and brain tumor localization~\cite{schutz2024framework}. However, it relies on preoperative static data, failing to account for soft tissue deformations and the complexities of tool-tissue interactions during surgery.

\subsection{Contributions} 
This paper presents an auditory-based AR tool designed to enhance nuanced decision-making in complex surgical tasks, addressing the limitations of MARS in augmented soft tissue deformation. It advocates for a systematic sonification approach that moves beyond discrete tissue states, focusing on defining tool-tissue interactions to (1) develop a generalizable model adaptable to diverse scenarios and (2) provide detailed insights that enable surgeons to integrate prior knowledge, explore the surgical field, deepen anatomical understanding, and minimize noise and errors. The aim is to deliver perceptually intuitive feedback that enhances cognitive and procedural understanding while supporting decision-making.

As a proof of concept, we propose a methodology for dynamic modeling and sonification of tool-tissue interactions. The framework translates tissue particle displacements—reflecting tissue characteristics and interactions—into excitation forces within a physics-based sound model, enabling physics-driven auditory feedback. Biomechanical numerical simulations, specifically Finite Element Modeling (FEM), are employed to model tool-tissue interactions by probing the anatomical model across different regions and trajectories. These simulations enable the representation of diverse tool-tissue dynamics.
We validate the approach through experiments analyzing the correlation between biomechanical displacement and acoustic output, optimizing the sonification model. User studies assess its technical validity, perceptual accuracy, and clinical relevance, providing guidelines for fine-tuning the model across various scenarios.

\section{BioSonix Design Framework}
\begin{figure}
\includegraphics[width=\textwidth]{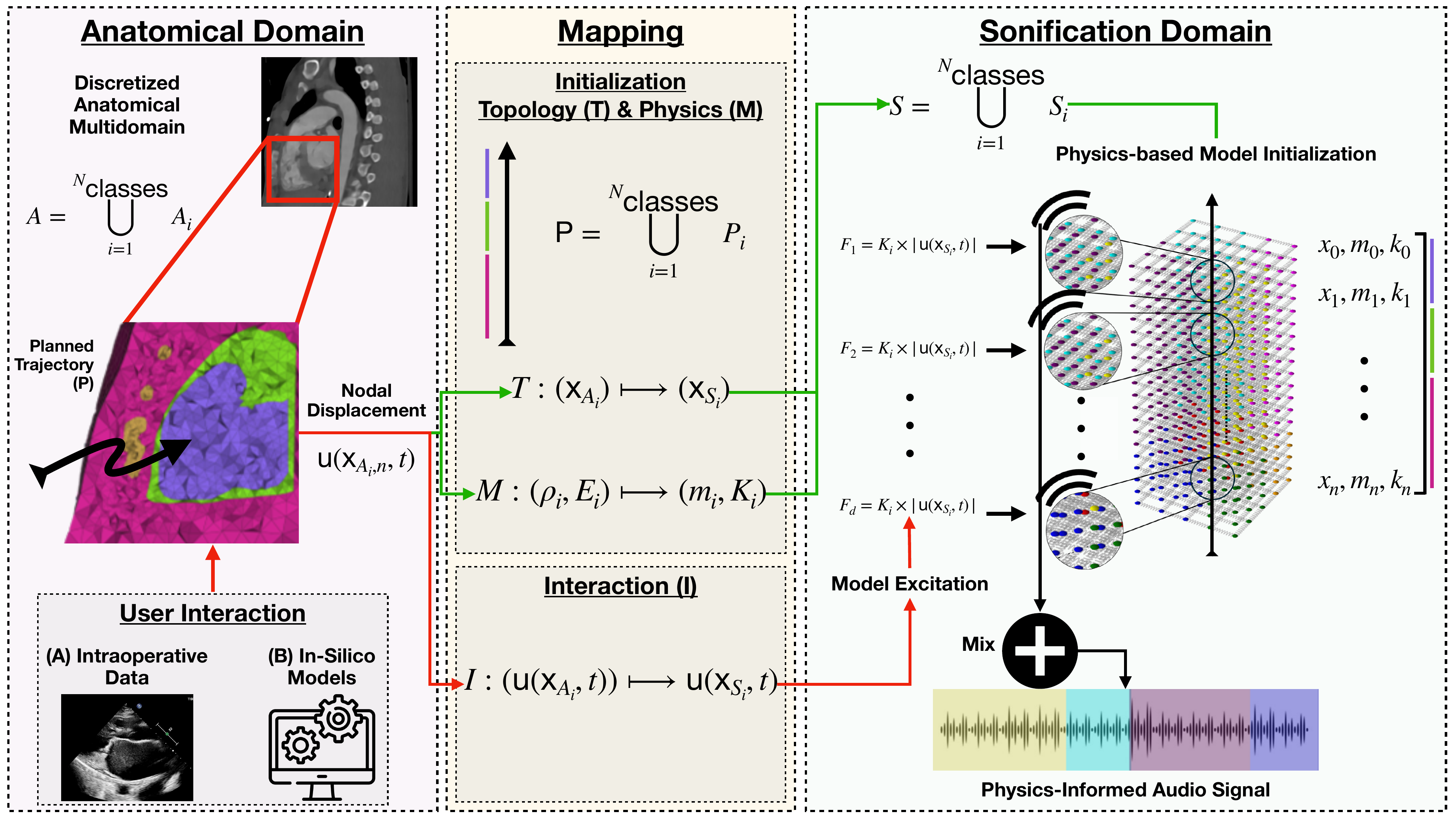}
\caption{BioSonix Design Framework -- A General Overview} \label{framework}
\end{figure}

The proposed framework, shown in Figure \ref{framework}, comprises three modules: (i) the Anatomical Domain Module, which reconstructs anatomy from pre-operative images while preserving tissue spatial information; (ii) the Mapping Module, which translates system physics into the Sonification Domain; and (iii) the Sonification Domain Module, which generates the final audio output. The following sections detail each module, their rationale, and interconnections.

\subsection{Anatomical Domain Module} 
Surgical tool navigation strategies are typically planned using preoperative static volumetric images, such as CT scans. These images enable 3D reconstruction of the anatomical region of interest (ROI) and the identification of safe pathways to target structures while avoiding critical tissues. Advanced processing software generates multidomain spatial discretizations of the ROI, preserving tissue classes and defining an anatomical discrete domain, $A$. 

The anatomical domain $A$ is represented as a union of subdomains, each corresponding to a specific tissue class. Mathematically, this can be expressed as:

\begin{equation}\label{eq}
A = \bigcup_{i=1}^{N_{\textit{classes}}} A_i, \quad \text{with} \quad A_i = \bigcup_{e=1}^{N_{\textit{el},i}} V_{e}^{(i)}.
\end{equation}

where  $A_i$  is the subdomain corresponding to tissue class $i$,  $N_{\textit{classes}}$ denotes the total number of tissue classes, $V_{e}^{(i)}$ represents a single volumetric element $e$ within subdomain $A_i$, and  $N_{\textit{el},i}$ is the number of volumetric elements in $A_i$. 

In the surgical field, the domain undergoes solicitations from sources such as physiological motions (e.g., heartbeat and respiration) and interactions between surgical tools and anatomical structures. These solicitations generate a 3D displacement field over time, which can be referred to the nodes of the domain $A$, and expressed as $\textbf{u}(\textbf{x}{A_i, n}, t),\quad i = 1, \dots, N{\textit{classes}}, \; n = 1, \dots, N_{\textit{nodes}}$, where  $\textbf{x}{A_i, n}$ are the spatial coordinates of the  n-th node in subdomain  $A_i$, and N{\textit{nodes}} is the total number of nodes in the anatomical domain $A$.

The displacement field at time $t$ is calculated as the difference in a node’s spatial coordinates between consecutive time frames

\begin{equation}
    \textbf{u}(\textbf{x}_{A_i, n}, t_i) = \textbf{x}_{A_i, n}(t_i) - \textbf{x}_{A_i, n}(t_{i-1}),
\end{equation}

or for simplicity, we will refer to the nodal displacement as $\textbf{u}(\textbf{x}_{A_i}, t)$, omitting the nodal index $n$ and focusing primarily on the tissue class $i$.

The temporal and spatial dynamics of the anatomical model can be derived through two distinct tracks, as shown in Fig.~\ref{framework}. In \textbf{Track A}, the displacement field $\textbf{u}(\textbf{x}{A_i}, t)$ is computed from intraoperatively acquired dynamic volumetric data, such as 3D ultrasound, ensuring high accuracy by capturing all motion sources. However, tool-tissue interactions in this track are deterministic. Conversely, \textbf{Track B} uses user-driven interactions to compute $\textbf{u}(\textbf{x}{A_i}, t)$, simulating dynamic behavior induced by surgical navigation through in-silico models that integrate system physics using computational methods.

\subsection{Mapping Module}
The Mapping Module defines the methodology for utilizing the system’s physics in $A$ to: (1) initialize and (2) excite the physics-driven sound model. (1) Initialization: The sound model is initialized using two mapping functions, $M$ and $T$. The function $M$ maps the physical properties:

\begin{equation}
M: (\rho_i, E_i) \longmapsto (m_i, K_i).
\end{equation}

Each subdomain $A_i$ is defined by specific density values $\rho_i$ and Young’s Modulus $E_i$, which approximate homogenized tissue properties and linear elastic mechanical behavior. The mapping function $M_i$ converts these values into mass $m$ and stiffness $K_i$ for the mass-spring model underlying the sound domain’s physical representation. To represent the spatial distribution of tissue classes, a new coordinate system \textbf{\textit{p}} is defined along the planned safe trajectory. By projecting the intersected elements $V_{e}^{(i)}$ of the 3D anatomical domain $A$ onto \textbf{\textit{p}}, the domain is reduced to a 1D coordinate system. This system is divided into segments labeled by tissue classes (i), where each segment’s length represents the distance traversed through that tissue. The 1D domain \textbf{\textit{P}} is defined as:

$\textbf{\textit{P}}: \bigcup_{i=1}^{N_{\textit{classes}}} \left[0, \ell_i\right]$,

where $\ell_i$ is the segment length for tissue class $i$. The mapping function $T$ then maps nodes from the Anatomical Domain $A$ to the corresponding nodes in the Sound Domain $S$ for each tissue class $i$:

\begin{equation}\label{eq:T}
T: (\textbf{x}_{A_i}) \longmapsto (\textbf{x}_{S_i}),
\end{equation}

2) The module for exciting forces ($F$) applied within the Sound Domain \(S\), in time and in space, is computed by the function $I$:

\begin{equation}\label{eq:I}
I: (\textbf{u}(\textbf{x}_{A_i}, t)) \longmapsto F(\textbf{x}_{S_i},t) = K_i\times|\textbf{u}(\textbf{x}_{S_i}, t)|.
\end{equation}

\subsection{Sonification Domain Module}\label{SonificationDomain}
The Sonification Domain Module is built on a physical model that implements mass-spring interactions, functioning as a virtual instrument initialized and excited by physical data. The domain $S$ is defined as:

$S = \bigcup_{i=1}^{N_{\textit{classes}}} S_i$,

where $S_i$ represents the subdomain corresponding to each tissue class $i$. Building on the sound physical model presented in ~\cite{villeneuve2019mass}, the defined coordinate system \textbf{\textit{p}} serves as the central axis, extending along 3D orthogonal directions to form a complex network of masses interconnected by linear spring elements.  The spatial distribution of tissue classes is propagated from \textbf{\textit{p}} to neighboring regions through the mapping function $T$, resulting into subsets of masses and springs specific to each tissue class. Using $M$, each mass subset is characterized by a mass value \(m_i\), and each spring subset is characterized by a stiffness value \(K_i\). The virtual instrument is excited by defining a number $d$ of \textit{Drivers} at multiple locations around the model (\(\textbf{x}_{S_i}\)). The excitation force $F(\textbf{x}_{S_i},t)$ applied at each driver is mapped by $I$. This results in multiple excited regions, generating specific wave patterns that encode the physical properties and dynamics of the anatomical domain. The signals generated from these excitation area are gathered by \textit{Listeners} and integrated into a Mixer, which produces the final audio output with a specific wave pattern encoding the physical properties and dynamics of the anatomical domain.
\section{Methods}\label{method-section} 
To establish a robust methodology for the proposed sonification framework, this study employs FEM simulations to model user interactions in a controlled and flexible environment. This section describes FEM-based simulations integrating user interactions into an abstract tissue model via needle insertions. These simulations capture dynamic tool-tissue interactions, facilitating the analysis of their correlation with resulting acoustic signals.

\begin{figure}[h!]
\centering
\includegraphics[width=\textwidth]{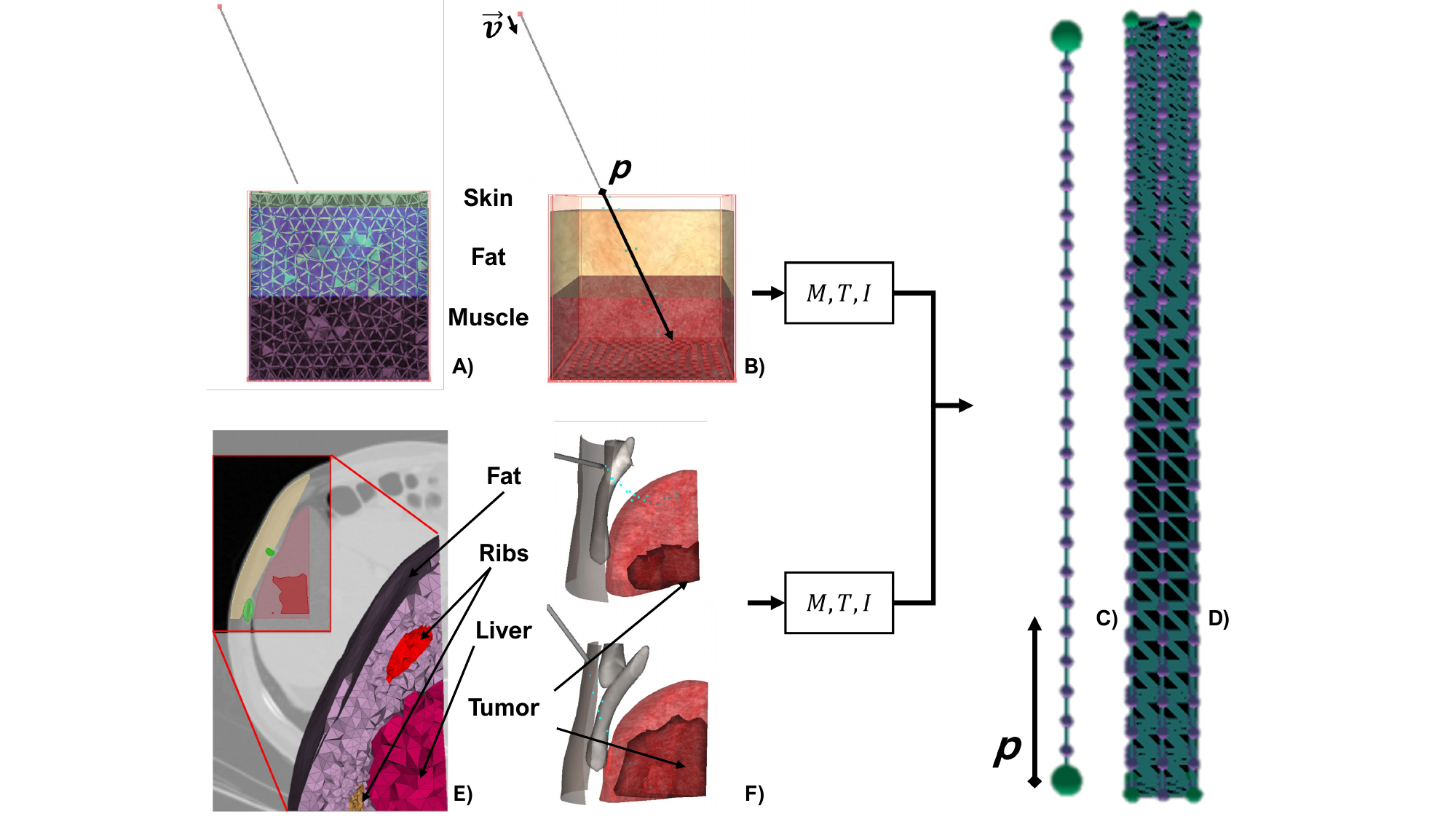}
\caption{FEM-based simulations to model tool-tissue dynamics by simulating insertion trajectories. (A)\textit{Anatomical Domain}: skin, fat, and muscle. (B) Tool insertion defined by an entry point, orientation \textbf{\textit{p}}, and a velocity $\vec{\boldsymbol{v}}$. (C) A 1D spring representing the basic sound model. (D) 3D advanced sound.}
\label{methodImg}
\end{figure}

\subsection{Model Implementation} \label{sound-model-section}
\textbf{Abstract Anatomical Model --} The simulations were conducted using the software SOFA Framework\footnote{\url{https://www.sofa-framework.org/}} with the Cosserat Plugin\footnote{\url{https://github.com/SofaDefrost/Cosserat}}, which is designed for the simulation of tool-tissue interactions.
The Cosserat Plugin includes an open-source implementation of needle insertion simulation, based on Cosserat Theory~\cite{Antman1995}. The interaction between the needle and tissue is modeled using the Complementary Constraints approach ~\cite{duriez2009needle}. Building upon this existing FEM simulation, an anatomical model was incorporated and defined as a cubic geometry ($25cm\times25cm\times25cm$), discretized into 13,465 tetrahedral elements. The bottom face of the cube is fixed, while a spring with a stiffness of $10^2 MPa$ is applied to the nodes on the lateral faces to prevent excessive lateral motion.
The non-homogeneous tissue model consisted of three distinct sub-domains, corresponding to three parallel layers (see Fig. \ref{methodImg} A), each representing a specific tissue class. In the presented preliminary study skin, fat and muscle are considered. The constitutive model of the tissue layers were approximated using a linear elastic constitutive law, with a co-rotational formulation employed to handle large deformations. Each sub-domain was characterized by its density ($\rho$) and a Young's Modulus ($E$), with values derived from literature ~\cite{mechanicalPropertiesReview}. A uniform Poisson's ratio of 0.4 was applied across all domains. 
The needle was modeled as a 17 Gauge, 26 cm-long rigid structure with a Young's modulus of 210 GPa. Each needle insertion was simulated by specifying an entry point on the top surface of the tissue model and an initial tool orientation in space. During the simulation, the needle was pushed from its handle with a constant velocity along the predefined trajectory. Herein, the simulated trajectory was defined as shown in Fig. \ref{methodImg} B and a velocity of $\vec{\boldsymbol{v}} = \frac{1 \text{ cm}}{s}$ is imposed. The remaining degrees of freedom of the tool were determined at each time step by solving the contact interaction with the tissue through the Complementary Constraint problem.

\begin{figure}[ht!]
\centering
\includegraphics[width=0.85\textwidth]{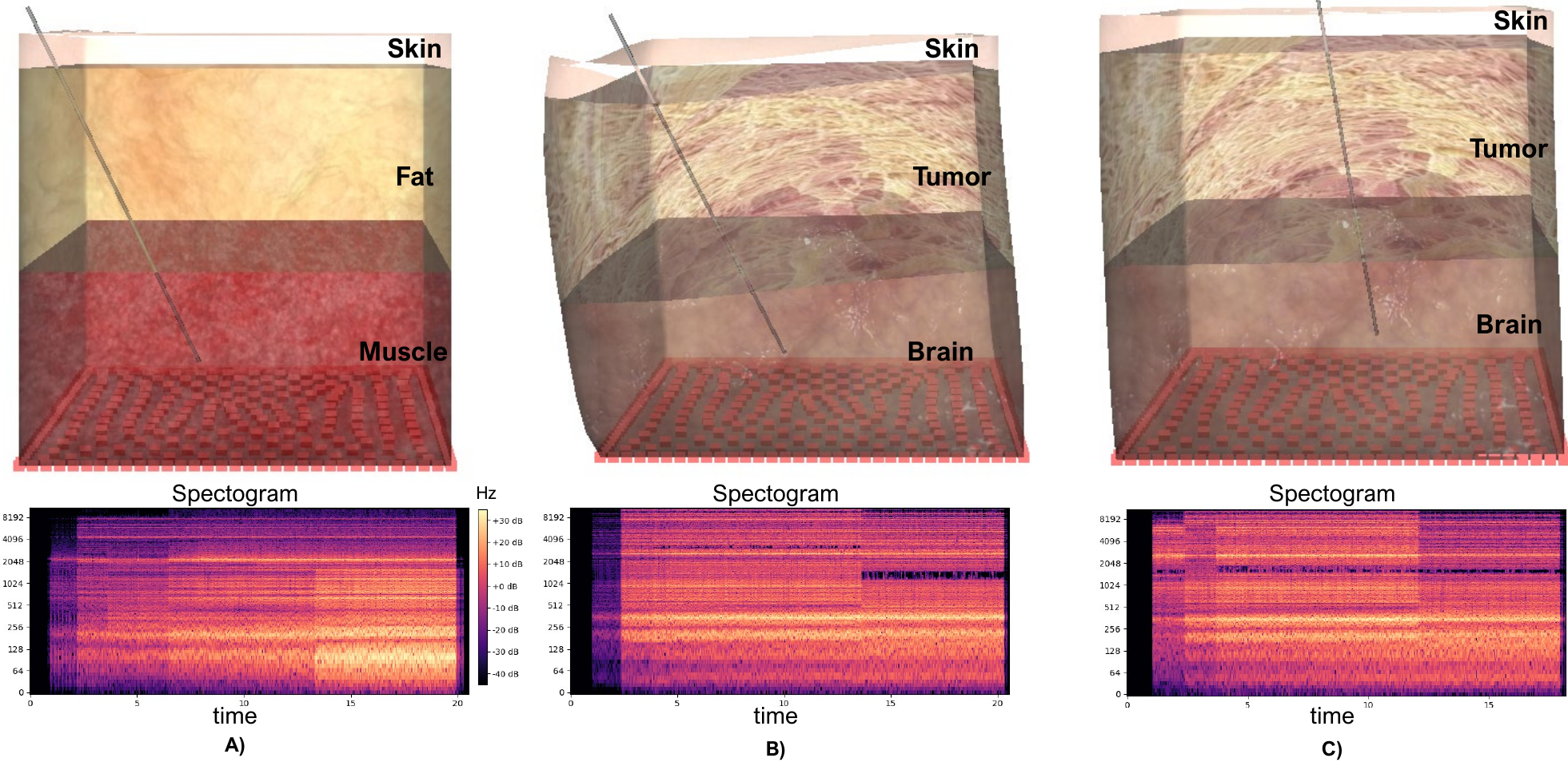}
\caption{Illustrative examples of the BioSonix design framework applied to abstract models: (A, B) Stability and responsiveness to tissue class variations; (B, C) Differentiation of dynamic tool-tissue interactions via sound signals.}
\label{abstract-models-img}
\end{figure}

\textbf{Toward Realistic Models --} Building upon the abstract model, new extensions of it were developed by: i)  incorporating additional tissue classes with varying thickness distributions within the multilayer cubic geometry; ii) implementing a new simulation setup to represent a tumor structure within a homogeneous medium; and iii) simulating diverse user interactions by varying trajectory definitions.
Fig. \ref{abstract-models-img} provides illustrative examples, and the detailed application of these models is described in Section \ref{user-study-1-aim}.

\textbf{Real Anatomical Models --} To demonstrate the applicability of the Bio\-Sonix design to real surgical tasks, the liver tumor biopsy procedure was considered. The modeling of the \textit{Anatomical Domain} $A$ is shown in Fig. \ref{methodImg} E and F. Starting from a public CT dataset \footnote{\url{https://github.com/wasserth/TotalSegmentator}}, a small region of interest (ROI) encompassing subcutaneous fat, liver, and ribs was segmented. The 3D reconstruction into a multi-domain model, reflecting distinct tissue classes, was built using the ANSA v24.0.1 software (Beta CAE Systems, USA). Four different tumor configurations were created. The application of this model is described in Section \ref{exp-evaluation-section}.

\textbf{Sound Model -- } The development of the Biosonix design framework was facilitated by the open-source miPhysics modelling libraries \footnote{\url{https://github.com/mi-creative/}}, which enabled the implementation of the physics-based sound model with the characteristics described in Section \ref{SonificationDomain}.
The \textbf{\textit{p}}-axis was defined as a straight path with the same orientation in space of the simulated trajectory traversing the whole model. 
The informative nodes of the anatomical model from wich the displacement was extracted ($\textbf{u}(\textbf{x}_{A_i}, t)$), were specific to the \textbf{\textit{p}}-axis and were mapped to the sound models by $T$ (light blue dots in Fig. \ref{methodImg} B)). The extracted $\textbf{u}(\textbf{x}_{A_i}, t)$, underwent minimal processing: the signals were first normalized within the $0.1$ to $0.9$ percentile range and subsequently scaled to a new range spanning from $0$ to $0.1$.
For the preliminary benchmark, the sound model was implemented in its simplest form: a 1D string directly projected from the \textbf{\textit{p}}-axis as shown in Fig. \ref{methodImg} C. The spring connections are 10 mm long.
The amount of \textit{Drivers} $d$ and \textit{Listeners}, and their location were defined with the method outlined in Section \ref{spatial-opt}.

\subsection{Model Optimization}\label{spatial-opt}
\begin{figure}[ht!]
\includegraphics[width=\textwidth]{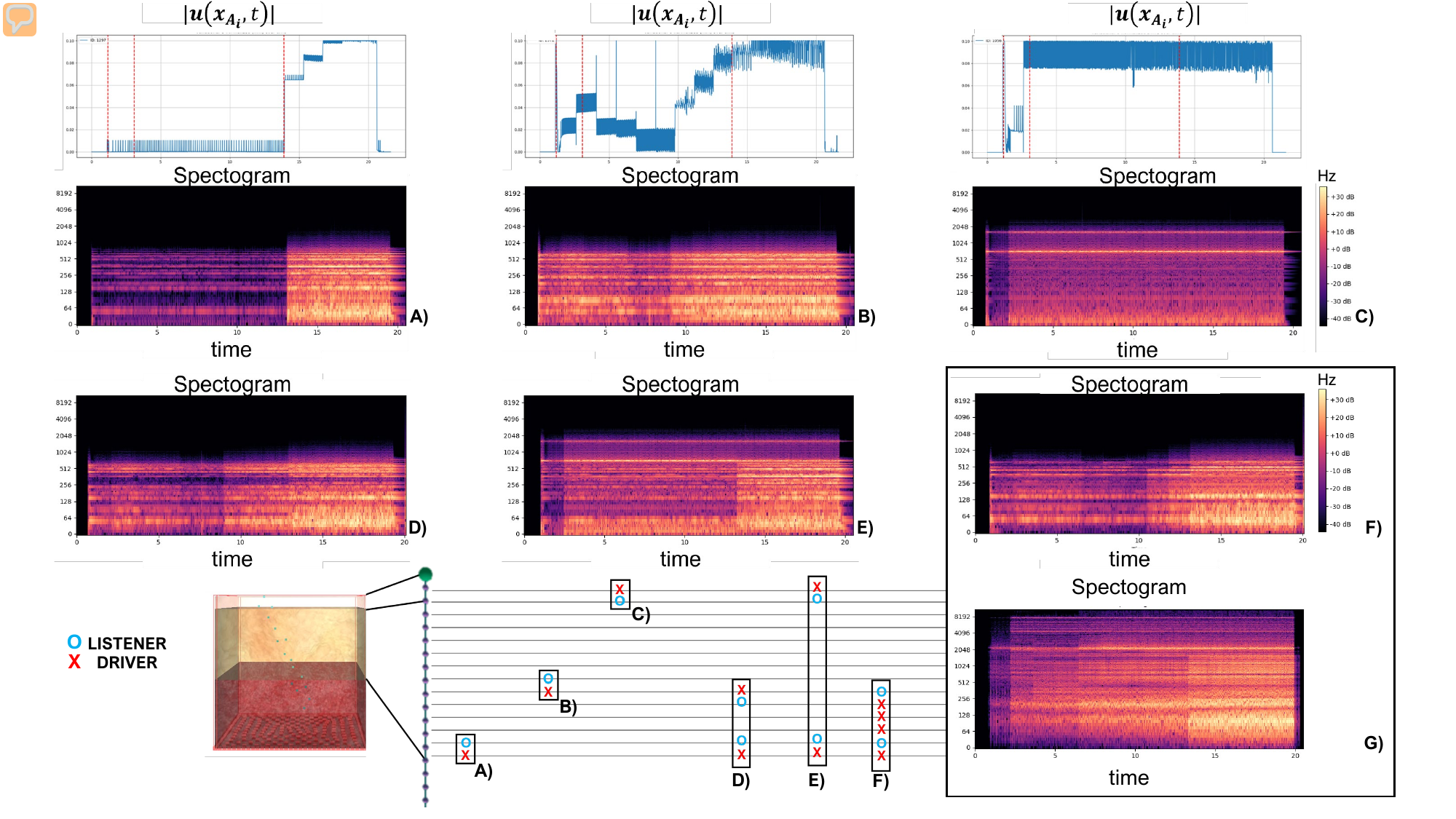}
\caption{Systematic experimentation to optimize the sound model: (A, B, C) \textit{Nodal Characterization} – correlation between displacement norms and resulting sounds; (D, E) \textit{Sonic Area Definition} – effects of varying \textit{Driver} distances; (F) \textit{Nodal Contribution Impact} – gradient from activating multiple nodes; (G) Fine-tuned model.}
\label{OptImg}
\end{figure}

To provide effective perceptual feedback to the user when touching the target anatomy, it was essential to define both $d$ and position of \textit{Drivers} starting from the target region, where the greatest emphasis is required. In this abstract model, muscle tissue was considered as the target structure. Within this controlled simulated environment, three systematic experiments were conducted to investigate the impact of the mapping functions (equations \ref{eq:T} and \ref{eq:I}) on the resulting sound.
In Fig. \ref{OptImg} few examples of this systematic approach are illustrated.

\textbf{Nodal Characterization --} Given that the physics-informed subsets of masses and springs were interconnected into a unified model, it was essential to analyze the contribution of each node to the resulting sound when excited. Fig. \ref{OptImg} A), B), C) illustrate the correlation between the norm of displacement data in the \textit{Anatomical Domain} and spectrograms. Notably, transitions corresponding to the tool puncturing the tissue—such as from skin to fat and from fat to muscle—are highlighted by red lines in the displacement plot.

\textbf{Sonic Area Definition --} The Sonic Area was defined as the minimal informative region above the target that contributes significant information to the resulting sound. Fig. \ref{OptImg} D and E illustrates the result of selecting a narrow area and a wider area.

\textbf{Nodal Contributions Impact --} With the Sonic Area defined, it was investigated how the final sound was influenced by adding contributing nodes from different locations within the Sonic Area. This aspect was particularly important for creating smooth sound transitions as the tool approached the target. Fig. \ref{OptImg} F) illustrates the resulting gradient in the spectrogram.

\textbf{Fine Tuning --} To further enhance the potential of the sound model, we introduced a more advanced design: a 3D topology formed by expanding the 1D structure along its orthogonal directions, centered on the \textbf{\textit{p}}-axis as shown in Fig. \ref{methodImg} D). Additionally, the function $M$ was empirically fine-tuned. The returned values $(m_i, K_i)$, were tested within the physics-based sound model to ensure stability and maximize perceptibility.
Fig. \ref{OptImg} G)  illustrates spectograms obtained by the fine-tuning of configuration Fig. \ref{OptImg} F).

\section{User Experiments and Expert Evaluation}
To evaluate the method’s user relevance and applicability to realistic tasks and clinical settings, we conducted two user studies. In all simulations, a constant insertion velocity of $\vec{\boldsymbol{v}} = \frac{1 \text{ cm}}{s}$ was applied, and minimal sound processing. Before each task, participants received brief training with three audiovisual examples of tool-tissue interactions(three for the general perceptibility study, and five for the expert evaluation). Each task then included 10 randomized audio-only examples.

\subsection{General Perceptibility Study} 
\label{user-study-1-aim}
The main objective of this study was to verify the perceptibility of sounds generated by the abstract models. A total of 41 audio samples were created using the optimized sound model and based on the abstract models (Fig. \ref{methodImg}A,B). Six soft tissue classes (fat, brain, muscle, skin, tumor), along with bone tissue to add variability.
The study involved three interactive tasks: \textbf{(1) Stopping:} detecting tool entry into critical tissue (always at the bottom), \textbf{(2) Tumor Detection:} identifying tool contact with a tumor, and \textbf{(3) Sorting:} sequencing tissue classes as the tool passed through them. We encourage readers to explore the audiovisual examples provided via this link: \url{https://tinyurl.com/47f2k8ku}.

\textbf{General Perceptibility Results --} A total of 22 participants (4 females, 20 males; aged 22–64) took part in the study, with diverse musical backgrounds ranging from minimal exposure to trained musicians. Most reported occasional or regular interaction with music but did not play instruments, ensuring a balanced perspective for auditory task analysis. Accuracy rates were 85.26\% for the sorting task (190 responses) and 83\% for the tumor detection task (200 cases). Statistics for the stopping task are provided in Table~\ref{tab:rotated_task_statistics}.

\begin{table}[h!]
\centering
\caption{Statistical results of the stopping task include Exact Match (\%)—the percentage of exact matches. High-performing values are bolded based on thresholds: Mean values close to 0 (-0.5 $\leq$ \text{Mean} $\leq$ +0.5), Std. Deviation below 1.0, and Exact Match (\%) above 70\%. Negative values indicate stopping early, positive values indicate stopping late (in seconds), and zero indicates precise stopping. T\# denotes the trajectory number, and the second value indicates the critical tissue layer thickness (in millimeters).}

\resizebox{\textwidth}{!}{
\begin{tabular}{|c|cc|cc|cc|}
\hline
\multirow{2}{*}{\textbf{Statistic}} & \multicolumn{2}{c|}{\textbf{Brain-Tumor}} & \multicolumn{2}{c|}{\textbf{Fat-Bone}} & \multicolumn{2}{c|}{\textbf{Fat-Muscle}} \\ \cline{2-7}
 & \textbf{T2, 80mm |} & \textbf{T1, 112mm} & \textbf{T1, 112mm |} & \textbf{T1, 160mm} & \textbf{T1, 112mm |} & \textbf{T1, 160mm} \\ \hline
\textbf{Mean [Sec]}               & -1.05 & 0.55  & 1.10  & -0.71 & -1.38 & 0.75  \\ \hline
\textbf{Std. Deviation}     & 1.19  & \textbf{0.51}  & 1.21  & \textbf{0.90}  & 3.76  & \textbf{0.72}  \\ \hline
\textbf{Median}             & -2.00 & 1.00  & 1.00  & 0.00  & 0.00  & 1.00  \\ \hline
\textbf{Exact Match}   & 20.00\% & 45.00\% & 15.00\% & 57.14\% & \textbf{76.19\%} & 10.00\% \\ \hline
\end{tabular}
}

\vspace{0.25em}

\resizebox{\textwidth}{!}{
\begin{tabular}{|c|cc|cc|cc|}
\hline
\multirow{2}{*}{\textbf{Statistic}} & \multicolumn{2}{c|}{\textbf{Muscle-Bone}} & \multicolumn{2}{c|}{\textbf{Muscle-Fat}} & \multicolumn{2}{c|}{\textbf{Tumor-Brain}} \\ \cline{2-7}
 & \textbf{T2, 112mm |} & \textbf{T1, 112mm} & \textbf{T2, 112mm |} & \textbf{T1, 180mm} & \textbf{T1, 112mm |} & \textbf{T1, 180mm} \\ \hline
\textbf{Mean [Sec]}               & -0.70 & 0.55  & -0.60 & \textbf{0.05}  & \textbf{-0.05} & \textbf{0.00}  \\ \hline
\textbf{Std. Deviation}     & \textbf{0.47}  & \textbf{0.51}  & \textbf{0.50}  & \textbf{0.40}  & \textbf{0.23}  & \textbf{0.20}  \\ \hline
\textbf{Median}             & -1.00 & 1.00  & -1.00 & 0.00  & 0.00  & 0.00  \\ \hline
\textbf{Exact Match}   & 30.00\% & 45.00\% & 40.00\% & \textbf{84.21\%} & \textbf{94.74\%} & \textbf{89.00\%} \\ \hline
\end{tabular}
}

\label{tab:rotated_task_statistics}
\end{table}

\subsection{Expert Evaluation of the Needle Biopsy Task}\label{exp-evaluation-section}
The simulation setup of Fig. \ref{methodImg} E and F was tested in this user study to assess the relevance of tasks to realistic medical scenarios, particularly the \textbf{Tumor Detection} task, which required identifying whether the tool touched the tumor. Twelve samples were generated by combining four tumor configurations with three trajectories. The second user study involved two male participants: a 41-year-old neuroradiologist with limited needle biopsy experience and a 46-year-old cardiologist who performs it weekly. Both had musical backgrounds as casual listeners but no experience playing instruments. Of the 20 cases, 70\% were correctly identified.

\section{Discussion and Conclusion}
BioSonix introduced an optimized method for modeling dynamic tool-tissue interactions with auditory feedback. Experimental evaluations demonstrated its robustness in capturing the behavior of tissue types with similar properties, by leveraging inherent tissue characteristics rather than artificial mappings. The system proved resilient to variations in interaction angles and insertion points, providing granular insights into tool-tissue dynamics. Machine-based evaluations, including spectral analysis, validated the model’s stability and reliable psychoacoustic properties, confirming its generalizability across scenarios. An initial user study assessed the effectiveness and perceptibility of sonifications in semi-interactive tasks. Designed for non-clinical users, this step laid the groundwork for future studies involving realistic surgical scenarios. The preliminary expert study demonstrated the method’s ability to represent realistic anatomical interactions with minimal training. While both user studies showed relatively high accuracy, participants noted a need for greater contrast between tissue classes, indicating low confidence despite accurate results. Improving tissue distinguishability, though not the primary focus here, remains an area for future exploration. Although methods such as~\cite{hermann2019data,weger2019real} could optimize model parameterization or enhance output signals~\cite{schwarz2007corpus,schwarz2012navigating}, this study was essential for identifying the key parameters relevant to such optimizations. The experiments revealed the model’s complexity and the influence of its parameters on the resulting sound. 

The mean time deviation in the stopping task was $\pm$ 1 second, equating to a 10 mm tool pathway distance within a $25cm\times25cm\times25cm$ tissue sample. This result is specific to the given tissue configuration and requires further testing to generalize across different dimensions or setups. The best performance occurred in the “Tumor-Brain” case, likely due to the higher contrast and increased thickness, though detailed studies are needed to confirm these factors. As sonification remains a relatively novel modality for interactive tasks, further research is needed to fully explore and realize its potential.


\subsubsection{\ackname} This work was supported by MUSA – Multilayered Urban Sustainability Action – project, funded by the European Union – NextGener- ationEU, under the National Recovery and Resilience Plan (NRRP) Mission 4 Component 2 Investment Line 1.5: Strenghtening of research structures and creation of R\&D “innovation ecosystems”, set up of “territorial leaders in R\&D”.

%

\bibliographystyle{splncs04}
\bibliography{mybib.bib}

\end{document}